# W4Λ: Leveraging Λ Coupled-Cluster for Accurate Computational Thermochemistry Approaches

Emmanouil Semidalas, Amir Karton, and Jan M. L. Martin*

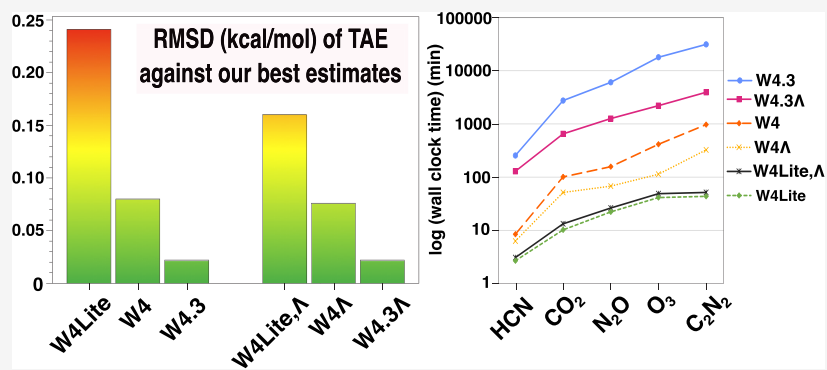

**ABSTRACT:** High-accuracy composite wave function methods like Weizmann-4 (W4) theory, high-accuracy extrapolated ab initio thermochemistry (HEAT), and the Feller−Peterson−Dixon (FPD) approach enable sub-kJ/mol accuracy in gas-phase thermochemical properties. Their biggest computational bottleneck is the evaluation of the valence post-CCSD(T) correction term. We demonstrate here, for the W4-17 thermochemistry benchmark and subsets thereof, that the Λ coupled-cluster expansion converges more rapidly and smoothly than the regular coupled-cluster series. By means of CCSDT(Q)$_\Lambda$ and CCSDTQ(5)$_\Lambda$, we can considerably (up to an order of magnitude) accelerate W4- and W4.3-type calculations without loss in accuracy, leading to the W4Λ and W4.3Λ computational thermochemistry protocols.

## 1. INTRODUCTION

Chemical thermodynamics and thermochemical kinetics are not just cornerstones of chemistry but arguably its very foundations. As the evaluation of absolute energies of molecules is a Sisyphian task (see Section 5 of ref 1 for a detailed discussion), the most fundamental thermochemical property of a molecule is generally taken to be the heat of formation. While this cannot be directly evaluated computationally, through the heats of formation of the gas-phase atoms, it can be related to the molecular total atomization energy (TAE)—the energy required to break up a molecule into its separate ground-state atoms. This latter quantity—a "cognate" of the heat of formation, if the reader permits a linguistic metaphor—is amenable to computation.

Experimental and theoretical thermochemical techniques have been recently reviewed by Ruscic and Bross.[2] Nowadays, by far the most reliable source of experimental (or hybrid) reference data are the Active Thermochemical Tables (ATcT) database,[3] in which a thermochemical network of reaction energies is jointly (rather than sequentially) solved[4,5] for the unknown heats of formation viz. TAEs and their respective uncertainties.

For small molecules, TAEs can now be evaluated by composite wave function [ab initio] theory (cWFT) approaches to a kJ/mol accuracy or better. Such approaches include Weizmann-4 (W4)[6,7] and variants (such as W4-F12,[8,9] W3X-L,[10] and Wn-P34[11]), the high-accuracy extrapolated ab initio thermochemistry (HEAT) by an international consortium centered on Stanton,[12−15] the Feller−Peterson−Dixon (FPD) approach,[16,17] which is less a specific cWFT than a general strategy, and the like. These approaches have been extensively validated against ATcT and other information: see, e.g., Karton[18] for a recent review.

The "Gold Standard of Quantum Chemistry" (T. H. Dunning, Jr.), CCSD(T) (coupled-cluster with all single and double substitutions with a quasiperturbative correction for triple substitutions),[19,20] performs much better than it has any right to, owing to a felicitous error compensation amply documented in refs 6, 7, and 12−15 (see Stanton[21] for a different perspective why this occurs). Post-CCSD(T) valence correlation corrections are the essential component that sets apart W4, HEAT, and the like from lower-accuracy approaches such as the correlation consistent composite approach (ccCA) by the Wilson group,[22] Gaussian-4 (G4) theory,[23,24] our own minimally empirical variants of these,[25,26] or indeed Weizmann-1 (W1) and W2 theory[27] and their explicitly correlated versions.[28] In all but the smallest cases, its evaluation is the single greatest "bottleneck" in W4 and HEAT calculations, owing to the extremely steep CPU time scaling of higher-order coupled-cluster approaches. [For fully iterative coupled-cluster up to m-fold connected excitations, the CPU time asymptotically scales as $n^m N_{virt}^{m+2} \times N_{iter}$, where $N_{iter}$ is the number of iterations, $n$ is the number of electrons correlated, and $N_{virt}$ is the number of virtual (unoccupied) orbitals. For quasiperturbative approximations, the corresponding scaling is $n^{m-1} N_{virt}^{m+1} N_{iter}$ for the underlying CC(m − 1) iterations and $n^m N_{virt}^{m+1}$ for the final step].

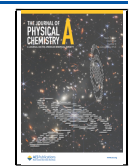

**Table 1. Overview of Post-CCSD(T) Methods and Basis Sets Used for Molecules in the W4-17 Database**

| method | basis set | #species | data set[a] |
|---|---|---|---|
| CCSDT | VDZ(p,s) through VQZ(f,d) | 200 | W4-17 |
|  | VQZ(g,d) | 199 | W4-17 except n-pentane |
|  | V5Z(h,f) | 65 | "W4.3" subset |
| CCSDT(Q) | VDZ(p,s) | 200 | W4-17 |
|  | VDZ(d,s) | 200 | W4-17 |
|  | VTZ(d,p) | 199 | W4-17 except $C_2Cl_6$ |
|  | VTZ(f,p) | 198 | W4-17 except $C_2X_6$ (X = F, Cl) |
|  | VTZ(f,d) | 197 | W4-17 except n-pentane, $C_2X_6$ (X = F, Cl) |
|  | VQZ(d,p) | 137 | W4-11 |
|  | VQZ(f,d) | 132 | W4-11: excluding 5 species[b] |
|  | VQZ(g,d) | 122 | W4-11: excluding 12 species, W4-08: without 3[c] |
|  | V5Z(h,f) | 59 | "W4.3" subset excluding 6 species[d] |
| CCSDT(Q)$_\Lambda$ | VDZ(p,s) | 200 | W4-17 |
|  | VDZ(d,s) | 200 | W4-17 |
|  | VTZ(d,p) | 188 | W4-17: excluding 12 species[e] |
|  | VTZ(f,p) | 157 | W4-11 plus 20 species from W4-17 |
|  | VQZ(g,d) | 122 | W4-11: excluding 12 species, W4-08: without 3[f] |
| CCSDTQ | VDZ(p,s) | 200 | W4-17 |
|  | VDZ(d,s) | 184 | W4-17: excluding 16 species |
|  | VTZ(d,p) | 134 | W4-08 except $AlF_3$, $AlCl_3$, $BF_3$, $O_2F_2$, $S_4$, $SO_3$ plus 25 species from W4-11 and 19 from W4-17 |
|  | VTZ(f,p) | 65 | "W4.3" subset |
|  | VQZ(g,d) | 50 | "W4.3" subset except 15 species[g] |
| CCSDTQ(5) | VDZ(p,s) | 193 | W4-17 except 7 species[h] |
|  | VDZ(d,s) | 165 | W4-17 except 32 species; W4-11 except $CH_3COOH$, $CF_4$, and $SiF_4$ |
|  | VTZ(f,p) | 53 | "W4.3" subset except 12 species |
| CCSDTQ(5)$_\Lambda$ | VDZ(p,s) | 193 | W4-17 except 7 species[h] |
|  | VDZ(d,s) | 163 | W4-17 except 34 species; W4-11 except $CH_3COOH$, $CF_4$, and $SiF_4$ |
|  | VTZ(f,p) | 53 | "W4.3" subset except 12 species |
| CCSDTQ5 | VDZ(p,s) | 96 | W4-08 |
|  | VDZ(d,s) | 65 | "W4.3" subset |
| CCSDTQ5(6) | VDZ(p,s) | 95 | W4-08 except $O_2F_2$ |
| CCSDTQ5(6)$_\Lambda$ | VDZ(p,s) | 95 | W4-08 except $O_2F_2$ |
| CCSDTQ56 | VDZ(p,s) | 88 | W4-08 except $BF_3$, $C_2N_2$, $O_2F_2$, $AlF_3$, $P_4$, $SO_3$, $S_4$, $AlCl_3$ |
|  | core–valence post-CCSD(T) correlation contributions | | |
| CCSDT − CCSD(T) | CVTZ(f,p) | 136 | W4-11 except cis-HOOO |
| CCSDT(Q) − CCSDT | CVTZ(f,p) | 118 | W4-08: excluding 8 species; W4-11: except 11 species[i] |

[a]W4-17, W4-11, and W4-08 data sets include 200, 137, and 96 species, respectively; the "W4.3" subset for which W4.3 results were obtained in earlier work contains 65 species. [b]Without acetic acid, ethanol, $CF_4$, $C_2H_5F$, and propane. [c]Excluding from W4-08: $FO_2$, $O_2F_2$, and ClOO; excluding from W4-11: acetaldehyde, formic and acetic acids, ethanol, glyoxal, cis-HOOO, $CF_4$, $SiF_4$, $C_2H_5F$, propane, propene, and propyne. [d]Without $CH_2CH$, $CH_2NH$, $NO_2$, $N_2O$, $H_2O_2$, and $F_2O$. [e]Without n-pentane, benzene, $C_2X_6$ (X = F, Cl), $PF_5$, $SF_6$, cis-$C_2F_2Cl_2$, cyclopentadiene, beta-lactim, $ClF_5$, n-butane, and allyl. [f]Excluding from W4-11: acetaldehyde, formic and acetic acids, ethanol, glyoxal, cis-HOOO, $CF_4$, $SiF_4$, $C_2H_5F$, propane, propene, and propyne; excluding from W4-08: $FO_2$, $O_2F_2$, and ClOO. [g]Without $CH_2C$, $CH_2CH$, $C_2H_4$, $CH_2NH$, HCO, $H_2CO$, $CO_2$, $NO_2$, $N_2O$, $O_3$, HOO, $H_2O_2$, $F_2O$, SSH, and HOF. [h]Without n-pentane, $C_2X_6$ (X = F, Cl), cyclopentadiene, silole, beta-lactim, and $ClF_5$. [i]Excluding from W4-08: $AlCl_3$, $AlF_3$, $BF_3$, $S_4$, $S_3$, $CS_2$, $SO_3$, and $P_4$; excluding from W4-11: allene, propyne, propane, $F_2CO$, $C_2F_2$, $C_2H_5F$, $SiF_4$, $CF_4$, cis-HOOO, glyoxal, and acetic acid.

Hence, any way to significantly reduce their computational cost or make their scaling less steep would extend the applicability of W4- and HEAT-type approaches.

As quasiperturbative triples, (T), proved so successful in ground-state coupled-cluster theory, attempts were then made to add them to excited-state equation-of-motion coupled-cluster theory with all singles and doubles (EOM-CCSD). This led also for the ground state to the so-called Λ coupled-cluster methods,[29−32] which recently seemed to show promise for computational thermochemistry as well.[33,34] Moreover, a recent study[35] on spectroscopic properties of small molecules likewise appeared to show that the Λ quasiperturbative series—CCSD(T)$_\Lambda$, CCSDT(Q)$_\Lambda$, CCSDTQ(5)$_\Lambda$, ...—converges more rapidly than the ordinary quasiperturbative expansion CCSD(T), CCSDT(Q), CCSDTQ(5), ... There was even a tantalizing hint in ref 35 that e.g., CCSDT(Q)$_\Lambda$ might be superior to CCSDTQ owing to a similar error compensation as one sees in CCSD(T) vs CCSDT.

This of course calls for a broader thermochemical exploration: we offer one in the present paper, focusing on the W4-17 benchmark[36] of 200 first- and second-row molecules, its W4-11 subset[37] published 6 years earlier, and the latter's W4-08 subset.[38] We shall show not only that Λ coupled-cluster indeed accelerates convergence but also that it can be exploited, with no loss in accuracy, for faster and less resource-intensive variants of W4 and W4.3 theory.

## 2. COMPUTATIONAL METHODS

All calculations were carried out on the Faculty of Chemistry's HPC facility ChemFarm at the Weizmann Institute of Science.

Geometries of the W4-17 set of molecules, originally optimized at the CCSD(T)/cc-pV(Q+d)Z level with only valence correlation included, were taken from the electronic supporting information (ESI) of the W4-17 paper[36] and used as-is, without further optimization.

Most of the post-CCSD(T) electronic structure calculations, and all of the post-CCSDTQ calculations, were carried out using the arbitrary-order coupled-cluster code[39−42] in the MRCC program system of Kállay and co-workers.[43] The specific levels of theory considered include CCSDT,[44] CCSDT[Q],[45] CCSDT(Q),[39] CCSDT(Q)$_A$,[41] CCSDT(Q)$_B$,[41] CCSDTQ,[46] CCSDTQ(5),[41] and CCSDTQ5.[47]

Coupled-cluster jobs were run in a "sequential restart" fashion where, e.g., CCSDT takes initial $T_1$ and $T_2$ amplitudes from the converged CCSD calculation, CCSDTQ in turn uses the converged CCSDT amplitudes as initial guesses for the $T_1$, $T_2$, and $T_3$ amplitudes, and so forth. For the open-shell species, unrestricted Hartree−Fock references were used throughout, except that higher-order triple excitation contributions, $T_3$ − (T), were also evaluated restricted open-shell in semicanonical orbitals as per the original W4 protocol.

The most demanding CCSDT, CCSDT(Q), CCSDT(Q)$_\Lambda$, and CCSDTQ calculations were carried out using a prerelease version of the NCC program developed by Matthews and co-workers as part of CFOUR.[48]

Basis sets considered are the cc-pVnZ basis sets of Dunning and co-workers[49,50] or truncations thereof. The abbreviated notation we use for truncated basis sets is probably best illustrated by example: VDZ(p,s) refers to cc-pVDZ truncated at p functions for nonhydrogen and at s functions for hydrogen, VDZ(d,s) refers to the untruncated cc-pVDZ basis set on nonhydrogen atoms, and the p polarization functions on hydrogen are removed.

It is well-known (e.g., refs 51 and 52) that for second-row atoms in high oxidation states, tight (i.e., high-exponent) d functions are energetically highly important at the CCSD(T), or even the Hartree−Fock (!), level. (This was ultimately rationalized[53] chemically as back-donation from chalcogen and halogen lone pairs into the vacant 3d orbital, which drops closer to the valence orbitals in energy as the oxidation state increases. A similar phenomenon involving tight f functions and vacant 4f and 5f orbitals exists in heavy p-block compounds.[54]) However, do tight d functions significantly affect post-CCSD(T) contributions? One of us[55] considered this question and found the total contribution to be quite modest and to largely cancel between higher-order triples and connected quadruples.

## 3. RESULTS AND DISCUSSION

### 3.1. Higher-Order Connected Sextuples. 
Table 1 summarizes our results using coupled-cluster methods for the molecular sets considered in this study.

The smallest contribution we will consider here are the higher-order connected sextuple excitations, CCSDTQ56 − CCSDTQ5(6) or $\hat{T}_6 - (6)$ for short, and CCSDTQ56 − CCSDTQ5(6)$_\Lambda$ or $\hat{T}_6 - (6)_\Lambda$ for short. A box-and-whiskers plot of these contributions is given in Figure 1. It can be seen there that $\hat{T}_6 - (6)_\Lambda$ has an extremely narrow spread, and that the largest outlier by far is C$_2$ at just 0.015 kcal/mol. We can hence consider CCSDTQ5(6)$_\Lambda$ to be essentially equivalent to CCSDTQ56 in quality.

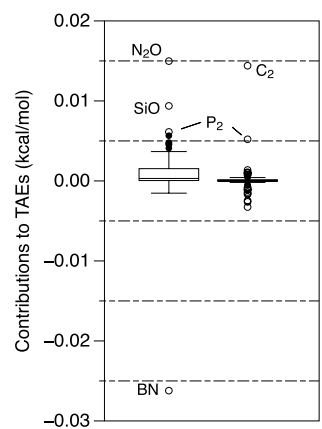

**Figure 1.** Box-and-whiskers plots of $\hat{T}_6 - (6)$ (on the left) and $\hat{T}_6 - (6)_\Lambda$ (on the right) using the VDZ(p,s) basis set for the W4-08 data set (excluding BF$_3$, C$_2$N$_2$, O$_2$F$_2$, AlF$_3$, P$_4$, SO$_3$, S$_4$, and AlCl$_3$). In this and subsequent box plots, the box boundaries represent the 25th and 75th percentile of the data, and the whiskers extend to the last point within 1.5 times the interquartile range (IQR) from the box, following the standard Tukey definition. Outliers are shown as filled circles if located more than 1.5 IQR from the box edge, while extreme outliers are represented by open circles when positioned more than 3 IQR from the box edge.

For CCSDTQ56 − CCSDTQ5(6), the spread is still very narrow, but now we have some small positive and negative outliers, BN −0.026, N$_2$O +0.015, SiO +0.009, P$_2$ +0.006 kcal/mol.

Since these tiny contributions are much smaller than the basis set incompleteness error in the larger components (see below), we feel justified in neglecting higher-order connected sextuples altogether.

### 3.2. CCSDTQ5(6)$_\Lambda$ − CCSDTQ(5)$_\Lambda$. 
The quasiperturbative sextuple contributions were evaluated only for the VDZ(p,s) basis set. A box plot can be seen in Figure 2. Whiskers are at +0.02 and −0.01 kcal/mol, around a median of basically 0.00 kcal/mol. The largest positive and negative outliers are +0.04 and −0.03 kcal/mol, respectively. In almost all situations, this contribution can be safely neglected.

If we used full iterative CCSDTQ56 − CCSDTQ5 instead (at massively greater computational expense), we would get

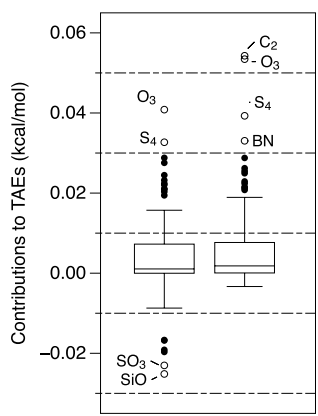

**Figure 2.** Box-and-whiskers plots of $(6)_\Lambda - (5)_\Lambda$ (on the left) and $(6)_\Lambda - T_5$ (on the right) using the VDZ(p,s) basis set for the W4-08 data set.

binding contributions throughout (small as they might be), topping out at 0.069 kcal/mol for $C_2$.

We argue that this contribution can be omitted in all but the most accurate work.

This is not the case for CCSDTQ5(6) − CCSDTQ(5), given the well-known deficiencies of CCSDTQ(5).[7,40]

**3.3. CCSDTQ(5)$_\Lambda$ − CCSDT(Q)$_\Lambda$.** We now move on to the $\Lambda$ connected quintuples contribution, CCSDTQ(5)$_\Lambda$ − CCSDT(Q)$_\Lambda$. The largest basis set for which we were able to evaluate even a subset (53 species) was VTZ(f,p).

As can be seen in Figure 3, the box is centered near 0.0 while the whiskers are ±0.04 kcal/mol in the largest basis set,

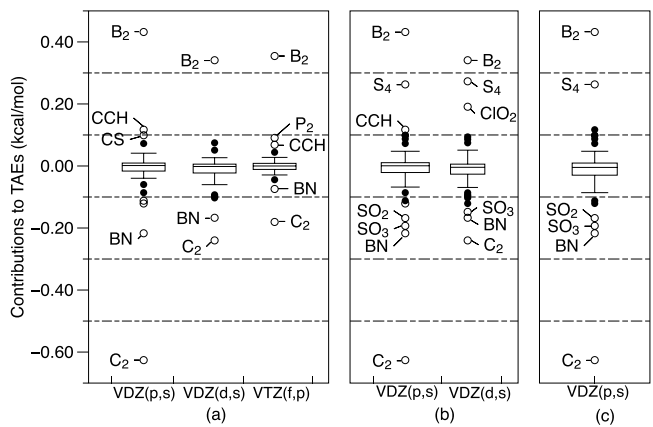

**Figure 3.** Box-and-whiskers plots of contributions to TAEs of $(5)_\Lambda - (Q)_\Lambda$ terms for (a) "W4.3" subset, (b) W4-08, and (c) W4-11 data sets.

VTZ(f,p). For the "W4.3" subset of 65 species, the root-mean square deviation (rmsd) between VDZ(p,s) and VTZ(f,p) is 0.06 kcal/mol, while that between VDZ(d,s) and VTZ(f,p) shrinks to just 0.02 kcal/mol.

For the smallest basis set, VDZ(p,s), there are sizable outliers, at −0.63 kcal/mol (for $C_2$) and +0.43 kcal/mol (for $B_2$). These shrink to −0.24 and +0.34 kcal/mol, respectively, for VDZ(d,s), i.e., upon expanding the nonhydrogen basis sets from double-$\zeta$ to polarized double-$\zeta$.

Obviously, for small "troublemakers" like $C_2$, BN, and $B_2$, switching to a basis set larger than VDZ(p,s) is not an issue at all.

In contrast, for CCSDTQ5 versus CCSDTQ, most contributions are positive. VDZ(p,s) has a box at about 0.08 kcal/mol with whiskers spanning 0.2 kcal/mol (outliers at 0.4 kcal/mol) while VDZ(d,s) has a smaller box (0.06 kcal/mol), and its whiskers span 0.14 kcal/mol (outliers up to 0.42 kcal/mol) (see Figure S1 in Supporting Information).

Therefore, aside from the much lower cost of CCSDTQ(5)$_\Lambda$ compared to CCSDTQ5, the CCSDTQ(5)$_\Lambda$ − CCSDT(Q)$_\Lambda$ contribution is close enough to zero that "in a pinch" it can be omitted altogether.

Nevertheless, let us also consider the quintuples contributions relative to CCSDTQ. For the 53 VTZ(f,p) species, the rms CCSDTQ(5)$_\Lambda$ − CCSDTQ is just 0.08 kcal/mol with the VTZ(f,p) basis set. For the same contribution, the rmsd between VDZ(p,s) and VTZ(f,p) basis sets is just 0.02 kcal/mol, while that between VDZ(d,s) and VTZ(f,p) shrinks to just 0.01 kcal/mol. (Box plots of CCSDTQ(5)$_\Lambda$ − CCSDTQ contributions to TAEs are shown in Figure S2 of Supporting Information).

What about (5) vs (5)$_\Lambda$ contributions compared to full iterative CCSDTQ5 − CCSDTQ? We have [CCSDTQ5 − CCSDTQ]/VDZ(d,s), or if you like, $\hat{T}_5$/VDZ(d,s) available for the "W4.3" subset of molecules, with which the rmsd of (5) is 0.05 kcal/mol, compared to just 0.01 kcal/mol for (5)$_\Lambda$. For the small VDZ(p,s) basis set, the rmsd between (5) and $\hat{T}_5$ is 0.04 kcal/mol, compared to 0.01 kcal/mol between (5)$_\Lambda$ and $\hat{T}_5$. We believe that this adequately shows the superiority of (5)$_\Lambda$.

**3.4. Higher-Order Quadruples.** The relevant statistics for the higher-order quadruples, $\hat{T}_4 - (Q)$ and $\hat{T}_4 - (Q)_\Lambda$, are given in Table 2, complemented by two box plots as shown in

**Table 2.** rmsd and Mean Signed Deviations (MSDs) in kcal/mol of $\hat{T}_4 - (Q)$ and $\hat{T}_4 - (Q)_\Lambda$ Terms for the "W4.3" Subset

| basis set | $\hat{T}_4 - (Q)$[a] | | $\hat{T}_4 - (Q)_\Lambda$[b] | |
|---|---|---|---|---|
| | rmsd | MSD | rmsd | MSD |
| neglecting | 0.255 | 0.109 | 0.111 | 0.044 |
| VDZ(p,s) | 0.094 | 0.039 | 0.062 | 0.007 |
| VDZ(d,s) | 0.030 | 0.011 | 0.014 | 0.001 |
| VTZ(d,p) | 0.015 | 0.003 | 0.010 | 0.001 |
| VTZ(f,p) | 0.012 | 0.006 | 0.007 | 0.005 |
| VQZ(g,d) | REF | REF | REF | REF |

[a]$\hat{T}_4 - (Q)$/VQZ(g,d) is used as reference. [b]$\hat{T}_4 - (Q)_\Lambda$/VQZ(g,d) is used as reference. A total of 50 data points are used in all comparisons.

Figures S3 and S4 while the outliers are listed in Table S1 in Supporting Information. It can be seen in Table 2 that complete neglect beyond CCSDT(Q) would cause an error of 0.255 kcal/mol, but only 0.111 kcal/mol beyond CCSDT-(Q)$_\Lambda$. Introducing a low-cost CCSDTQ/VDZ(p,s) calculation would reduce rmsd to 0.094 kcal/mol for $\hat{T}_4 - (Q)$ but 0.062 kcal/mol for $\hat{T}_4 - (Q)_\Lambda$—the latter is small enough that one is tempted to substitute a single CCSDTQ(5)$_\Lambda$/cc-pVDZ(p,s) calculation for the W4 theory combination of CCSDTQ/cc-pVDZ and CCSDTQ5/cc-pVDZ(p,s).

Similarly, with the cc-pVDZ(d,s) basis set, we have rmsd = 0.030 kcal/mol for $\hat{T}_4 - (Q)$, but just 0.014 kcal/mol for $\hat{T}_4 - (Q)_\Lambda$. With $\hat{T}_4 - (Q)$, one needs to escalate to VTZ(f,p) (like in W4.3 theory) to achieve a comparable rmsd = 0.012 kcal/mol.

**3.5. $(Q)_\Lambda - (Q)$.** The difference between ordinary parenthetical quadruples and their $\Lambda$ counterpart approaches zero for systems dominated by dynamical correlation but becomes quite significant when there is strong static correlation. [In ref 6, we defined the %TAE[(T)] diagnostic, the percentage of the CCSD(T) TAE that is due to connected triples, as a "pragmatic" diagnostic for static correlation (see also refs 56 and 57 for other diagnostics)]. Between %TAE[(T)] and %TAE[$(Q)_\Lambda - (Q)$], and upon eliminating the irksome BN diatomic, the coefficient of determination $R^2$ for 160 closed-shell species is 0.7164 with the cc-pVDZ(d,s) basis set, which increases to 0.7407 upon additionally eliminating $ClF_5$. While this is not something one would want to substitute for an actual evaluation, it does indicate a relationship between the two quantities.

Compared to the largest basis set for which we have sufficient data points available, namely, VQZ(g,d), the rmsd is 0.066 kcal/mol for VDZ(p,s) but drops to 0.027 kcal/mol for VDZ(d,s) and to 0.01 kcal/mol or less for a VTZ basis set.

In W4 theory, we combined[6] CCSDT(Q)/VTZ(f,d) with [CCSDTQ − CCSDT(Q)]/VDZ(d,p). If, for the sake of argument, we split up the latter term into [CCSDT(Q)$_\Lambda$ − CCSDT(Q)]/VDZ(d,p) plus [CCSDTQ − CCSDT(Q)$_\Lambda$]/VDZ(d,p), then based on the previous section, we could prune the basis set for the second step to VDZ(p,s) and greatly reduce computational expense.

Continuing this line of argument, in the W4.3 and W4.4 theories, CCSDT(Q)/V{T,Q}Z extrapolation is combined with [CCSDTQ-CCSDT(Q)]/VTZ(f,d). If we again partition the latter term into a relatively cheap [CCSDT(Q)$_\Lambda$ − CCSDT(Q)]/VTZ(f,p) and a very expensive [CCSDTQ − CCSDT(Q)$_\Lambda$]/VTZ(f,p) step, we could again take down the basis set for the latter to VDZ(d,s).

Why not do the largest basis set $(Q)_\Lambda$ to begin with? The extra computational expense of evaluating the "left eigenvector" is of course one factor but not the main one: in practice, we find the additional memory requirements to be a greater impediment for large molecules and basis sets.

**3.6. Parenthetical Connected Quadruples (Q).** In W4lite and W4 theory, the (Q) contribution is included via scaling, as it was shown[6] that extrapolation of (Q) from too small basis sets yields erratic results for highly polar molecules like $H_2O$ and HF.

For calibration, we used V{Q,5}Z extrapolation for 58 species ("W4.3" subset minus $CH_2CH$, $CH_2NH$, $NO_2$, $N_2O$, $H_2O_2$, and $F_2O$). Minimizing rmsd against that (see Table 3), we find the cheapest option that still has a tolerably small rmsd to be VDZ(d,s) scaled by 1.227, close enough to the 5:4 used in the past for $\hat{T}_4$ in W3 theory.[58] The error drops to 0.082 kcal/mol with 1.17 × VTZ(d,p) and to 0.038 kcal/mol with 1.099 × VTZ(f,p), only semantically different from 1.1 × VTZ used in W4 theory. This is roughly half the rmsd of V{D,T}Z extrapolation, at rmsd = 0.074 kcal/mol (Figure S5).

V{T,Q}Z as used for W4.3 theory in ref 6, at rmsd = 0.006 kcal/mol is essentially as accurate as the reference. Importantly, however, deleting the highest angular momentum in both basis sets is found to cause only negligible further loss of accuracy, to rmsd = 0.01 kcal/mol. This offers an attractive way to reduce the cost of W4.3 calculations (see below).

Table 3. rms (kcal/mol) of CCSDT(Q)-CCSDT Errors for TAEs in the "W4.3" Subset

| basis set | rmsd |
|---|---|
| VDZ(p,s) | 0.413 |
| VDZ(d,s) | 0.249 |
| VTZ(d,p) | 0.183 |
| VTZ(f,p) | 0.109 |
| VTZ(f,d) | 0.108 |
| VQZ(f,d) | 0.058 |
| VQZ(d,p) | 0.149 |
| VQZ(g,d) | 0.044 |
| V5Z(h,f) | 0.021 |
| 1.227 × VDZ(d,s)[a] | 0.139 |
| 1.170 × VTZ(d,p)[a] | 0.082 |
| 1.099 × VTZ(f,p)[a] | 0.038 |
| V{D(d,s),T(f,p)}Z[b] | 0.074 |
| V{T(d,p),Q(f,d)}Z[a] | 0.010 |
| V{T(f,p),Q(g,d)}Z[b] | 0.006 |
| (Q)/V{Q(g,d),5(h,f)}Z | REF |

[a]Scaling factor obtained from rmsd minimization. [b]Extrapolation exponent a = 2.85 for V{D,T}Z and a = 3.25 for V{T,Q}Z; no change occurs with Karton's extrapolation exponents (a = 2.9968 and a = 3.3831) from Table 5 in ref 59. A total of 58 species are included in all comparisons.

**3.7. Higher-Order Connected Triples, CCSDT − CCSD(T).** For a subset of 65 molecules within W4-08—the so-called "W4.3" subset for which we were able to perform W4.3 calculations in refs 36 and 37—we managed to carry out CCSDT/cc-pV5Z calculations, and hence, we used CCSD-(T)/V{Q,5}Z extrapolation as the reference value. These limits are apparently not very sensitive to the extrapolation exponent, as there is just 0.007 kcal/mol rms difference between values obtained using a fixed 3.0 and Karton's optimized value[59] of 2.7342. That means that the reference we are using is pretty insensitive to details of the extrapolation procedure; hence it makes sense to calibrate the $\hat{T}_3 - (T)$ higher-order connected triples contribution to TAE by comparison with [CCSDT − CCSD(T)]/V{Q,5}Z extrapolations.

The rms $\hat{T}_3 - (T)$ contribution is 0.684 kcal/mol. It is obvious from Figure 4 and Table 4 that untruncated and truncated cc-pVDZ basis sets are barely better than doing nothing, with rms errors over 0.5 kcal/mol. A cc-pVTZ basis set recovers a more respectable chunk, but still leaves 0.18 kcal/mol rmsd, which increases to 0.25 kcal/mol if the f functions are omitted. In order to get below 0.1 kcal/mol without extrapolation, at least a cc-pVQZ basis set is required, although the g function can apparently be safely omitted. V5Z reaches 0.05 kcal/mol.

Extrapolation from cc-pV{D,T}Z, as practiced in W4 theory,[6] brings down rmsd to just 0.041 kcal/mol with the exponent 2.7174 optimized by Karton:[59] if one substitutes the 2.5 recommended in refs 6 and 7, rmsd slightly increases to 0.052 kcal/mol. For cc-pV{T,Q}Z, Karton's exponent is functionally equivalent to 2.5, and we obtain rmsd = 0.012 kcal/mol—that level as used in W4.3[6] and W4.4[7] can credibly be used as a reference. Alas, unlike for (Q), removal of the top angular momenta of both basis sets increases the error to the same as cc-pV{D,T}Z, presumably primarily because of the impact on cc-pVTZ.

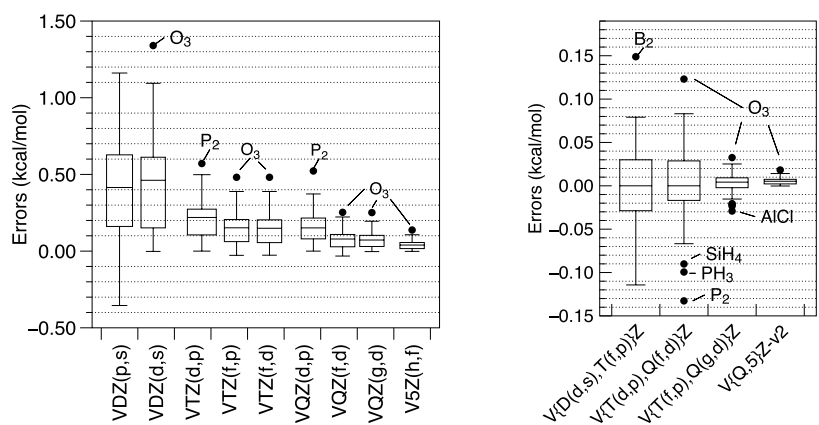

**Figure 4.** Errors for $\hat{T}_3 - (T)$ terms vs $\hat{T}_3 - (T)/V\{Q(g,d),5(h,f)\}Z$ for the "W4.3" subset with 65 molecules.

**Table 4. rmsd $\hat{T}_3 - (T)$ against $\hat{T}_3 - (T)/V\{Q,5\}Z$ (kcal/mol) for the "W4.3" Subset**

| basis set | rmsd[a] |
|---|---|
| neglecting | 0.684 |
| VDZ(p,s) | 0.554 |
| VDZ(d,s) | 0.545 |
| VTZ(d,p) | 0.248 |
| VTZ(f,d) | 0.179 |
| VTZ(f,p) | 0.181 |
| VQZ(f,d) | 0.094 |
| VQZ(g,d) | 0.090 |
| VQZ(d,p) | 0.194 |
| V5Z(h,f) | 0.049 |
| Schwenke coefficient $A_L$[b] | rmsd[a] |
| V{D,T}Z | 0.503 | 0.041 |
| V{D,T}Z | 0.498[c] | 0.041 |
| V{D,T}Z | 0.570 | 0.052 |
| V{T,Q}Z | 0.950 | 0.012 |
| V{T(d,p),Q(f,d)}Z | 0.528[d] | 0.042 |
| V{Q,5}Z-v2 | 1.049 | 0.007 |
| V{Q,5}Z | 1.210 | REF |

[a]All 65 molecules of the "W4.3" subset are considered for all comparisons. [b]The energy is extrapolated with the Schwenke[60] two-point formula $E_\infty = E + A_L(E(L) - E(L-1))$. [c]Exponent taken from Karton.[59] [d]Obtained from rmsd minimization.

### 3.8. Post-CCSD(T) Core−Valence Contributions.

Core−valence $\hat{T}_3 - (T)$ is required for W4.2 and W4.3 theory[6] while core−valence (Q) corrections enter in W4.4 theory.[7] In the original papers,[6,7] we employed the cc-pCVTZ basis set;[61] in the present work, we employed the combination of cc-pCVTZ on nonhydrogen atoms and cc-pVTZ(no d) on hydrogen—abbreviated, CVTZ(f,p). Full CCSDT proved feasible for all of W4-11 (except for cis-HOOO, owing to an SCF convergence issue) while (Q) was feasible for W4-08 minus eight species (seven of which contain multiple second-row atoms with their [Ne] cores) and for the additional W4-11 species minus 10 larger first-row species and SiF$_4$.

The rms core−valence $\hat{T}_3 - (T)$ contribution was 0.046 kcal/mol and the rmsd core−valence (Q) term was 0.037 kcal/mol—larger than the remaining errors in the valence post-CCSD(T) part of W4.3 theory and hence not negligible.

W4.2 theory is identical to W4 theory except for the core−valence $\hat{T}_3 - (T)$ term. As discussed in ref 6, it removes the dependence on the specific definition of frozen-core CCSD(T) for restricted open-shell references (semicanonicalization before transform as in Gaussian,[62] MRCC, and CFOUR versus transform before semicanonicalization as in MOLPRO[63]).

### 3.9. Composite Approaches. 

First, let us concentrate on W4 itself. In composite A, we retain the $\hat{T}_3 - (T)$ component from original W4, except that we remove the top angular momentum from hydrogen: CCSDT/{VDZ(d,s),VTZ(f,p)}. Next, we add (Q)/VTZ(f,p) and scale the contribution by 1.11. Then, we add $[(Q)_\Lambda - (Q)]/VDZ(d,s)$ and $CCSDTQ(5)_\Lambda/VDZ(p,s) - CCSDT(Q)_\Lambda/VDZ(p,s)$. In effect, we replace the CCSDTQ/VDZ(d,s) and CCSDTQ5/VDZ(p,s) steps with a single CCSDTQ(5)$_\Lambda$/VDZ(p,s) step.

Composite A is feasible for the entire W4-17 data set except for seven species where the quintuples present an obstacle. As they are expected to be of minor importance in these species, one can substitute CCSDTQ/VDZ(p,s) as a fallback option, leaving us with a complete W4-17 set. By way of illustration, for NCCN (dicyanogen) on 16 cores of an Intel Ice Lake server at 2.20 GHz with 768 GB RAM and local SSD, the CCSDT(Q), CCSDTQ, and CCSDTQ5 steps of standard W4 theory take 16.2 h wall time, compared to 5.4 h for the corresponding steps in Composite A, making it 3 times faster. We thus renamed composite A to W4Λ. The rmsd between this W4Λ and standard W4 is 0.066 kcal/mol for the post-CCSD(T) contributions to TAEs of the W4-08 data set.

Now, let us consider W4.3. In composite B, we retain the $\hat{T}_3 - (T)$ component from original W4, except that we remove the top angular momentum from hydrogen: CCSDT/{VTZ(f,p),VQZ(g,d)}. The (Q) we extrapolate from (Q)/{VTZ(d,p),VQZ(f,d)}. Then, we add $[(Q)_\Lambda - (Q)]/VTZ(f,p)$ and $CCSDTQ(5)_\Lambda/VDZ(d,s) - CCSDT(Q)_\Lambda/VDZ(d,s)$.

This composite B is feasible for nearly all of the W4-11 set; in conjunction with the CCSDT/CVTZ(f,p) core−valence contribution, we dub it here W4.3Λ. The rmsd between W4.3Λ and standard W4.3 is only 0.019 kcal/mol for the TAEs of 65 species in the "W4.3" data set.

In composite C, we add a CCSDTQ5(6)$_\Lambda$/VDZ(p,s) − CCSDTQ(5)$_\Lambda$/VDZ(p,s) component to composite B. Together with the CCSDT(Q)/CVTZ(f,p) core−valence contribution, we propose this as W4.4Λ.

What about lower-cost options? A form of W4lite,Λ can be created by scaling (Q)$_\Lambda$/VDZ(d,s) by an empirical scale factor obtained by minimizing the rmsd from W4.3Λ: we thus find a scale factor of 1.249 (in practice, 5:4) and rmsd = 0.149 kcal/mol.

At a reviewer's request, the final leaf of the ESI workbook compares the original W4-17 atomization energies at W4 and W4.n levels with their Λ counterparts. As the differences between them are essentially confined to the post-CCSDT terms (leaving aside the small differences in $T_3 − (T)$ owing to the present omission of the highest angular momentum in the hydrogen basis set), differences between the two are very small as expected, with an IQR = 0.04 kcal/mol. The one glaring exception is FOOF, where the discrepancy reached nearly 1 kcal/mol. Upon close scrutiny, we were able to identify an error in the published W4-11 and W4-17 values for that molecule and to trace those to a CCSDT(Q)/cc-pVTZ calculation where a malfunctioning restart from saved amplitudes (back in 2007) yielded an erroneous total energy of −349.4117061 $E_h$ rather than the correct value of −349.4131206 $E_h$. As a result, the published[36] $TAE_e$ value of 151.00 kcal/mol should read 151.89 kcal/mol, and its $TAE_0$ counterpart of 146.00 should read 146.89 kcal/mol instead. In addition, in Table S2 of the article mentioned above, the $T_4$ contribution for FOOF should read 3.18 rather than 2.29 kcal/mol.

### 3.10. Timing Comparison of Composite Approaches.

The total electronic energy of a W4-type composite approach is the sum of CCSD(T) energy near the basis set limit and various post-CCSD(T) terms, detailed in previous sections. As for all but the smallest molecules, the cost of the post-CCSD(T) contributions dwarfs that of the CCSD(T)/CBS step, our timing comparison can focus exclusively on the former. For example, three separate single-point energy calculations are needed for W4: (a) CCSDTQ5/VDZ(p,s); (b) CCSDTQ/VDZ(d,s); and (c) CCSDT(Q)/VTZ(f,p). The $(\hat{T}_3 − T)/V\{D,T\}Z$ term is a byproduct of parts (b) and (c). Note that CCSDT(Q), CCSDTQ5, and CCSDTQ56 scale with system size as $N^9$, $N^{12}$, and $N^{14}$, respectively.[18]

Figure 5 shows total wall clock times (note the logarithmic y axis) for calculating the post-CCSD(T) terms of several composite approaches across five species in the W4-08 data set. Calculations were performed on 16 cores and in 340 GB of RAM on otherwise empty nodes with dual 26-core Intel Ice Lake CPUs at 2.2 GHz, 7.0 TB of local solid state disk, and 768 GB of RAM. Wall clock times and percentages of time elapsed per step are provided in Tables S2 and S4 of Supporting Information.

Λ-based composites, especially for larger molecules, exhibit significant speedups over non-Λ counterparts. For cyanogen, standard W4 takes 16.3 h, while W4Λ only requires one-third as much (5.4 h); W4.3 requires 522.3 h, but W4.3Λ only one-eighth thereof (65.6 h). Both W4lite and W4lite,Λ are almost of equivalent cost for the molecules considered. As can be seen from Table S4, for W4 the highest-order coupled-cluster step, CCSDTQ56/VDZ(p,s), will dominate the CPU time as the molecules grow larger; in contrast, for W4Λ, the CCSDT(Q)/VTZ step remains the dominant one. If the calculation is carried out wholly using MRCC, then the (Q) step benefits from almost perfect parallelism, which offers a further way to speed up the W4Λ calculation if enough cores are available. If, on the other hand, the CCSDT(Q) step is carried out using the very fast NCC module in CFOUR, then this will further favor W4Λ over W4 as the dominant iterative quintuples step in the latter is not amenable to NCC at present. For much larger systems, the total cost of canonical calculations would of course become intractable, but their CPU time scaling can be nearly linearized by substituting localized natural orbital coupled-cluster, LNO-CCSDT(Q),[64−66] for conventional CCSDT(Q). The performance of the latter is presently being investigated in our laboratory.

## 4. CONCLUSIONS

Using the W4-17 data set (and the W4-11 and W4-08 subsets thereof) as our "proving ground", we have reconsidered post-CCSD(T) corrections in computational thermochemistry in light of Λ coupled-cluster methods. It is apparent that the Λ approach converges more rapidly and smoothly with the substitution levels. Our findings corroborate our earlier conjecture[35] that the coupled-cluster series has two more "sweet spots" (performance-price optima) beyond CCSD(T), namely, CCSDT(Q)$_\Lambda$ and CCSDTQ(5)$_\Lambda$. These findings are then exploited to drastically reduce the computational requirements of the W4 and W4.3 computational thermochemistry protocols; we denote the modified versions W4Λ and W4.3Λ, respectively.

## ■ ASSOCIATED CONTENT

### Data Availability Statement

This material is also freely available from the Figshare repository at http://doi.org/10.6084/m9.figshare.24806913. Additional data can be obtained from the authors upon reasonable request.

### Ⓢ Supporting Information

The Supporting Information is available free of charge at https://pubs.acs.org/doi/10.1021/acs.jpca.3c08158.

> Total energies, TAE contributions, and statistics (ZIP)
>
> Box plots of CCSDTQ5 − CCSDTQ contributions to TAEs, CCSDTQ(5)$_\Lambda$ − CCSDTQ contributions, CCSDTQ − CCSDT(Q) and CCSDTQ − CCSDT(Q)$_\Lambda$ basis set convergence, CCSDT(Q) − CCSDT differences; relative errors of $\hat{T}_4 − (Q)$ and $\hat{T}_4 − (Q)_\Lambda$ compared to $[\hat{T}_4 − (Q)]/VQZ(g,d)$ and $[\hat{T}_4 − (Q)_\Lambda]/VQZ(g,d)$; wall clock times of composite approaches for selected species; wall clock times per post-CCSD(T) step of each composite approach; refs 43, 62, 63 with full author lists (PDF)

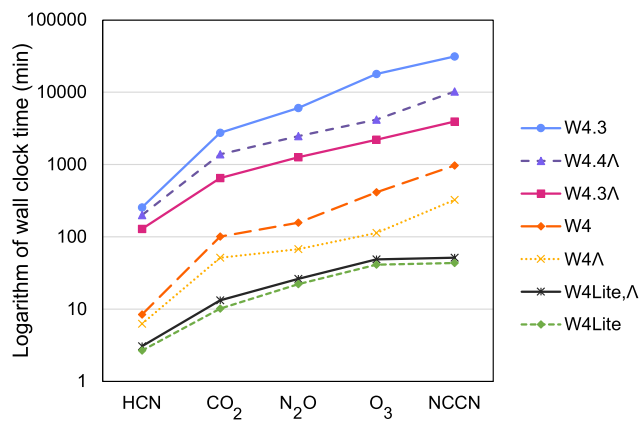

**Figure 5.** Wall clock time for calculating the post-CCSD(T) terms of various composite schemes for five molecules in the W4-11 data set. The y axis is in the logarithmic scale.

■ AUTHOR INFORMATION

**Corresponding Author**

Jan M. L. Martin − *Department of Molecular Chemistry and Materials Science, Weizmann Institute of Science, 7610001 Reḥovot, Israel;* orcid.org/0000-0002-0005-5074; Email: gershom@weizmann.ac.il

**Authors**

Emmanouil Semidalas − *Department of Molecular Chemistry and Materials Science, Weizmann Institute of Science, 7610001 Reḥovot, Israel;* orcid.org/0000-0002-4464-4057

Amir Karton − *School of Science and Technology, University of New England, Armidale, New South Wales 2351, Australia;* orcid.org/0000-0002-7981-508X

Complete contact information is available at:
https://pubs.acs.org/10.1021/acs.jpca.3c08158

**Notes**

The authors declare no competing financial interest.

■ ACKNOWLEDGMENTS

Research at Weizmann was supported by the Israel Science Foundation (grant 1969/20), by the Minerva Foundation (grant 2020/05), and by a research grant from the Artificial Intelligence for Smart Materials Research Fund, in memory of Dr. Uriel Arnon. E.S. thanks the Feinberg Graduate School (Weizmann Institute of Science) for a doctoral fellowship and the Onassis Foundation (Scholarship ID: FZP 052-2/2021-2022). J.M.L.M. is the incumbent of the Baroness Thatcher [of Kesteven] Professorial Chair in Chemistry. The authors would like to thank Drs. John F. Stanton, Devin A. Matthews, and Branko Ruscic for helpful discussion and Dr. Margarita Shepelenko for proofing the draft manuscript. J.M.L.M. would like to dedicate this paper to the memory of his untimely deceased colleague and friend, Prof. Abraham Minsky (1952−2024), Renaissance man and physical organic chemist turned structural biologist extraordinaire. May his memory be blessed.